\documentclass{pasj00}
\usepackage{psfig}
\draft

\begin{document}
\SetRunningHead{M. Mutoh et al.}{A New Test for the GPS Absorption Mechanism}
\Received{2001/10/12}
\Accepted{2001/12/22}

\title{A New Test for the Absorption Mechanism of GPS Radio Sources Using Polarization Properties}

\author{Mutsumi \textsc{Mutoh},\altaffilmark{1,2} Makoto \textsc{Inoue},\altaffilmark{2} Seiji \textsc{Kameno},\altaffilmark{2} Keiichi \textsc{Asada},\altaffilmark{1,2,3}\\
 Kenta \textsc{Fujisawa},\altaffilmark{2} and Yutaka \textsc{Uchida}\altaffilmark{1}
}

\altaffiltext{1}{Physics Department, Science University of Tokyo, Kagurazaka, sinjuku-ku, Tokyo 162-0833}
\email{mutou@hotaka.mtk.nao.ac.jp}
\altaffiltext{2}{National Astronomical Observatory, 2-21-1 Osawa, Mitaka, Tokyo 181-8588}
\altaffiltext{3}{Department of Astronomical Science, Sokendai, 2-21-1 Osawa, Mitaka, Tokyo 181-8588}


\maketitle

\begin{abstract}
We consider the use of polarization properties as a means to discriminate between Synchrotron Self-Absorption (SSA) and Free--Free Absorption (FFA) in GHz-Peaked Spectrum (GPS) sources.  The polarization position angle (PA) of synchrotron radiation at high frequencies for the optically thin regime is perpendicular to the magnetic field, whereas it is parallel to the magnetic field at low frequencies for the optically thick regime.  Therefore, SSA produces a change in PA of
$90^{\circ}$ across the spectral peak, while FFA does not result in such a change.  We analyzed polarization data from VLA observations for six GPS sources to see if such a change in PA was present.  Our results indicate that there is no significant evidence for $90^{\circ}$ change in PA across the spectral peak, suggesting that FFA is more likely than SSA for low-frequency cutoffs in these sources.
\end{abstract}

\section{Introduction}

GHz-Peaked Spectrum (GPS) sources are characterized by a simple convex radio spectrum with a peak at GHz frequencies. The common properties of bright samples of GPS sources are: small size ($<$ 1 kpc), powerful radio luminosity ($L>10^{45}$ erg s$^{-1}$), and apparently low variability (e.g., \cite{Stanghellini1998}; \cite{O'Dea1998}). GPS sources associated with quasars tend to show a complex morphology in VLBI images, while those associated with galaxies are likely to be double or triple radio sources (\cite{Phillip-Mutel1980}; \cite{O'Dea1991}).
The spectral shape of GPS sources is distinct from that of extended radio sources. Generally, extended radio galaxies have steep spectral profiles, in contrast to the flat spectral profiles of quasars.
GPS sources have steep spectra at high frequencies, as do radio galaxies, which indicates optically thin synchrotron radiation. However, the low-frequency cutoff in the spectral profile is peculiar to GPS sources. There has been a considerable debate regarding the interpretation of this low-frequency cutoff.

The low-frequency cutoff is ascribed by some to Synchrotron Self-Absorption (SSA).  SSA is usually discussed in terms of the equipartition brightness temperature, $T_{\rm B}$ (assuming equipartition of energy between the radiating particles and the magnetic field), which is derived from the flux density and frequency of the spectral peak, and the optically thin spectral index. \citet{Readhead1996} showed that the observed peak brightness temperature is consistent with the equipartition brightness temperature for the lobes of 2352+495.
\cite{Snellen2000} suggested SSA based on correlations between the peak frequency and the angular size, and between the peak flux density and the angular size.

On the other hand, \citet{Bicknell1997} suggested that the observed anticorrelation between $\nu_{\rm m}$ and size can be explained by Free--Free Absorption (FFA).  
VSOP and multi-frequency ground VLBI observations of OQ 208 have been 
used to infer that FFA is the more likely mechanism, with the two 
radio lobes surrounded by a cold dense plasma which causes thermal
FFA (\cite{Kameno2000}). Assuming intrinsically symmetric double
lobes, the apparent asymmetry of opacities towards
lobes was explained as being caused by the differences in path lengths 
through the plasma,
which has an estimated electron density and electron temperature of $ 600 <
n_{\rm e} < 7\times 10^{5}$ and $10^{4} < T_{\rm e} < 6\times 10^{7}$,
respectively.
In addition, FFA is favored for other
GPS sources, such as 0108+388 \citep{Marr2001} and NGC 1052
\citep{Kameno2001}.

So far, the origin of the low-frequency cutoff has been discussed in terms of the spectral properties and brightness temperatures. However, since both FFA and SSA can produce similar spectral shapes, it is difficult to differentiate between the two mechanisms in this way.
Here, we examine the possibility of discriminating FFA from SSA by using the polarization properties of sources. In section~2 we review the relevant polarization properties, and we describe the data analysis in section 3, followed by section 4 where the results for the sources studied to date are given.

\section{A New Test of Absorption Mechanisms Using Polarization Properties}

\citet{Aller1970} pointed out that the polarization position angle (PA) is perpendicular to the magnetic field when synchrotron emission is optically thin in terms of SSA, while PA is parallel to the magnetic field in optically thick regimes. This is explained as follows.
The ratio of the specific intensities, $I_{\rm B} / I_{\rm A}$, determines the PA, where $I_{\rm A}$ and $I_{\rm B}$ are specific intensities parallel and perpendicular to the magnetic field, respectively. The specific intensity is approximately proportional to the emissivity, $\epsilon$, when the emission is optically thin, and can be described as
\begin{eqnarray}
\frac{I_{\rm B}}{I_{\rm A}}\simeq\frac{\epsilon_{\rm B}}{\epsilon_{\rm A}}=\frac{3\gamma+5}{2}>1.
\end{eqnarray}
Here, we assume that the number density of relativistic electrons follows a power law,  $N(E) \: dE = K E^{-\gamma} \: dE$, as a function of energy $E$ where $K$ is a constant. When the emission is optically thick, the specific intensity is also inversely proportional to the absorption coefficient, $\kappa$, so that
\begin{equation}
\frac{I_{\rm B}}{I_{\rm A}}\simeq\frac{\epsilon_{\rm B}/\kappa_{\rm B}}{\epsilon_{\rm A}/\kappa_{\rm A}}=\frac{3\gamma +5}{3\gamma +8}<1.
\end{equation}
Comparing equations (1) and (2), we can see that PA changes by
90$^{\circ}$ between optically thick and optically thin regimes. The
optical depth of synchrotron radiation decreases with increasing
frequency, so that the radiation will be 
optically thick at low frequencies and
optically thin at high frequencies.
The spectral peak
occurs at the frequency where the optical depth is approximately
unity. Therefore, the 90$^{\circ}$ jump in PA appears across $\nu_{\rm m}$.

Such a $90^{\circ}$ jump of PA is found in some BL Lac objects. 
Absorption in BL~Lac objects is thought to be due to
SSA (\cite{Readhead1996}; \cite{Shen1999}).  Figure~1 shows (a) the total
flux densities and (b) the PAs of the BL~Lac object 0212+735 
over a period of 4 yr from observations at
the University of Michigan Radio Astronomy Observatory (UMRAO).
The PA at 14.5 GHz jumped by $90^{\circ}$ after MJD 40100,
while at 4.8~GHz it remained almost constant.
The total flux densities indicate that the emission was
optically thick between 4.8 and 14.5~GHz before MJD 40100, 
but became optically thin at 14.5 GHz. This variability is thought to be due to the expansion of a jet component.
This synchronous change of the opacity and PA suggests a causal relationship between them, implying SSA rather than FFA in BL Lac objects. 
Another example of a $90^{\circ}$ jump is reported for the BL~Lac object
OJ~287 \citep{Gabuzda2001}. They detected a $90^{\circ}$ jump in
PA for the core between 5~GHz and 22~GHz, and interpreted it as the 
transition from the optically thick to optically thin domain.

The above properties of PA can be used as a test of the
absorption mechanism in GPS sources. If the low-frequency cutoff is
caused by SSA, we expect to observe a $90^{\circ}$ jump in
PAs across the spectral peak frequency. Otherwise, in the
case of FFA, no such $90^{\circ}$ jump will be seen.
(This, of course, assumes that the emission from optically thick regime dominates the polarized emission, an assumption which is easily tested by observations with sufficient angular resolution to resolve the contributions to the polarized flux from the various source components.)
%
Based on polarization data just below $\nu_{\rm m}$, \citet{Stanghellini1998} mentioned the possibility that the polarized emission may not come from the region responsible for the optically thick emission, or that the turnover is not due to SSA.
%
Because GPS sources are generally not time variable, we combined multi-frequency observations around $\nu_{\rm m}$ from various epochs to investigate the existence of the $90^{\circ}$ jump.

\section{Data Analysis}
\subsection{The Samples from VLA Observations}

We analyzed the PAs of six GPS sources observed with the VLA
(\cite{Saikia1998}; \cite{Stanghellini1998}). Our selection criteria
were: 

(1) to have more than 2 polarization observations in both optically thick and thin frequency regimes,

(2) to have a percentage polarization higher than 0.1$\%$ at
5 GHz.

\noindent The sources 0248+430, 0711+356, 0738+313, 0743$-$006, 1143$-$245
and 2134+004 met these criteria. They have all been identified
as quasars, and their properties are summarized in table~1.  
These sources have between 4 and 6 polarization observations at
different frequencies. The PAs and percentage polarizations
are given in table 2, and the PAs of the six sources as a
function of squared wavelength ($\lambda^2$) are plotted in figure 2. 
The vertical lines in these plots indicate $\nu_{\rm m}$.

\subsection{Statistical Study}

We analyzed the distribution of the PAs statistically. In the case of FFA,
PA follows the Faraday Rotation formula across the spectral peak,
which can be fitted by
\begin{eqnarray}
<\theta_{\it k}>_{\rm FFA}&=&\theta_0+RM\lambda_{\it{k}}^2,
\end{eqnarray} 
at both optically thick and thin frequencies. Here, $\theta_{\it{k}}$
is the PA observed at $\lambda_{\it k}$, and $\lambda_{\rm m}$ is wavelength
corresponding to $\nu_{\rm m}$.  The free parameters $RM$ and
$\theta_0$ are the Faraday Rotation Measure and the intrinsic PA,
respectively.  In the case of SSA, PA is fitted as follows:

\begin{eqnarray}
<\theta_{\it k}>_{\rm SSA}=
\left\{
\begin{array}{lr}
\theta_0+RM\lambda_{\it k}^2 \pm 90^{\circ} & \mbox{$(\lambda_{\it k}>\lambda_{{\rm m}})$},\\
\theta_0+RM\lambda_{\it k}^2  & \mbox{$(\lambda_{\it k}<\lambda_{\rm m})$}.
\end{array}
\right.
\end{eqnarray}

The sign ``$\pm$'' indicates jump in PA between the optically thick
and thin regimes. Figure~3 shows an example of PAs for 1143$-$245,
with and without $\pm 90^{\circ}$ offsets for the data points above the 
peak frequency. We fit both the FFA and SSA models to the data and evaluated 
the residuals for each source. 
Here, we introduce "$dist(\theta)$" as the distance of PA from the fit,
\begin{eqnarray}
dist(\theta)&=&\frac{1}{2}(1-\cos2\theta)\\
&=&\sin^2\theta.
\end{eqnarray}
This function has a $180^{\circ}$ period, and is thus free from the
$180^{\circ}$ ambiguity in PA. When $\theta$ is sufficiently small, $dist(\theta_{\it k}-<\theta_{\it k}>)$ is approximately equal to
$(\theta_{\it k}-<\theta_{\it k}>)^2$.  Here, we define the residual, $\it{S}$, as
\begin{eqnarray}
S&=&\sum_{\it k}\frac{dist(\theta_{\it{k}}-<\theta_{\it{k}}>)}{\sigma_{\it{k}}^2}.
\end{eqnarray}
We use the ratio, $F_{\rm{rat}}$, defined as
\begin{eqnarray}
F_{\rm{rat}} &=& \frac{S_{\rm{SSA}}}{S_{\rm{FFA}}},
\end{eqnarray}
to determine which absorption mechanism is more likely.
Unless there is any significant difference of likelihood between the FFA
and SSA models, as a first-order hypothesis, the ratio,
$F_{\rm{rat}}$, would follow the F-distribution,
\begin{eqnarray}
F(\nu_{\rm SSA},\nu_{\rm FFA}) &=&\frac{\chi_{\rm SSA}^2/\nu_{\rm SSA}}{\chi_{\rm FFA}^2/\nu_{\rm FFA}},
\end{eqnarray}
where $\nu_{\rm SSA}$ and $\nu_{\rm FFA}$ are the numbers of degrees
of freedom corresponding to the $\chi^2$ values, $\chi_{\rm SSA}^2$ and
$\chi_{\rm FFA}^2$, respectively. The ratio of variances,
\begin{eqnarray}
\frac{\chi_{\nu_{\rm SSA}}^2}{\chi_{\nu_{\rm FFA}}^2} &=& \frac{S_{\rm SSA}}{S_{\rm FFA}},
\end{eqnarray}
is also distributed as $F_{\rm{rat}}$.  Evaluating $F_{\rm{rat}}$, 
one can thus discriminate between two models. Simply, if
$F_{\rm rat}>1$, then FFA is more likely.

\section{Results}

In table 3 and figure 4, we show the results of statistical
analysis. The value of $F_{\rm{rat}}$ is greater than 1 for all sources
except 2134+004. This implies that FFA is more likely than SSA.  For
0711+356, $F_{\rm{rat}} = 8.586$ indicates that the residual of the
FFA model is significantly less than that of the SSA model. This means
that a simple SSA model cannot be accepted as the cause of the
spectral cutoff below 1.6~GHz in this source. 
In the cases of 0738+313 and 0248+430,
$F_{\rm{rat}}$ = 2.431 and $F_{\rm{rat}}$ = 2.067, respectively,
again indicating that the FFA model is more likely.  For
1143$-$245, $F_{\rm{rat}} = 1.304$. This is because of the high $RM$ 
in this source, which results in small values of $S$ for both
models (see figure~3).  For 0743$-$006 and 2134+004, because we have only four 
data points, the degree of freedom is very small. Consequently, the values of $S$ for both SSA and
FFA are small, and $F_{\rm{rat}}$ is close to 1, as can be seen
in figure~4b.

\section{Discussion}

Although five of the six sources have $F_{\rm{rat}} > 1$, the case for FFA is not compelling, in part because of lack of observed frequencies, particularly around $\nu_{\rm m}$.
The SSA model predicts not only the $90^{\circ}$ jump in PA, but also the change in the fractional polarization as a function of the wavelength. In the optically thin regime, the degree of polarization decreases with decreasing frequency, and becomes null at $\nu_{\rm m}$ (\cite{Aller1970}).  It increases again at the optically thick regime, but the polarization degree is $\sim$ 1/10 of that at the optically thin regime.
The percentage polarization is plotted as a function of $\lambda^2$ for the six sources studied here in figure 5. The variation in the degree of polarization expected from the SSA model cannot be seen for any source except 0738+313. 
This also implies that SSA is unlikely to be the cause of the low-frequency cutoff.

For 2134+004, $F_{\rm rat}<$1 results from the low spatial resolution against the complex structure of the source components.  In fact, 2134+004 has two components, which have different $RM$s \citep{Taylor2000}.  In such a case, our method should be applied to the individual components.  The $RM$ asymmetry of 2134+004 is easily explained by means of the FFA model.  Suppose that two intrinsically identical lobes are surrounded by dense plasma to produce FFA. Because there is usually a difference in the path lengths to the lobes in the plasma, the $RM$ differences reflect the differences in the path length. Such a model allows us to estimate the electron density and the magnetic field \citep{Kameno2000}.

In conclusion, studies of the polarization properties of six published GPS source have yielded no convincing evidence for the $90^{\circ}$ change in PA across the spectral peak expected from SSA models. This and the distribution of fractional polarization as a function of wavelength lead us to conclude that FFA may be the dominant cause for the low-frequency cutoffs in most of these sources.

\bigskip
\bigskip
\bigskip
\bigskip
This research made use of data from the University of Michigan
Radio Astronomy Observatory, which is supported by funds from the
University of Michigan. We thank Philip G. Edwards for much advice.


\begin{table*}
\begin{center}
\caption{Sample of GPS sources.}
\vspace{6pt}
\begin{tabular}{cccccc}
\hline\hline
Source & {\it z} & {\it m} & $S_{5 GHz}$ (Jy) & $\nu_{\rm max}$ (GHz) & $S_{\rm max}$ (Jy)\\
\hline
 0248+430  & 1.316 & 15.5V & 1.24 & 5.2 & 1.27\\
 0711+356  & 1.620 & 17.0V & 0.861 & 1.6 & 1.42\\
 0738+313  & 0.631 & 16.1V & 3.66 & 5.1 & 3.82\\
 0743$-$006  & 0.994 & 17.5V &  2.05 & 6.0 & 2.12\\
 1143$-$245  & 1.950 & 18.5V & 1.40& 2.0 & 1.69\\
 2134$-$004  & 1.936 & 16.8V & 8.59& 5.2 & 8.59\\
\hline
\end{tabular}
\end{center}
\vspace{6pt}
\par\noindent
The columns are: source name, red shift, optical magnitude and filter/color, flux density at 5 GHz, observed peak frequency and peak flux density (Stanghellini et al. 1998).
\par\noindent
\end{table*}

\begin{table*}
\begin{center}
\caption{Polarization of GPS samples}
\vspace{6pt}
\begin{tabular}{r||rrrrrrrrrrrr}
\hline\hline
frequency &  1335& &  1665& &  4535 && 4985 & & 8085 & &8465&\\
source & $\%$ & PA & $\%$ & PA & $\%$ & PA & $\%$ & PA & $\%$ & PA & $\%$ & PA\\
\hline
 
0248+430  & 1.7 & +52 & 1.5 & +23 & 2.0 & $-$18 & 2.0 & $-$27 & 0.5 & $-$40 & 0.4 & $-$44\\

0743$-$006  & 0.4 & $-$76 & 0.5 & $-$7.0 & 0.2 &  & 0.2 & & 0.8 & +1.0 & 1.0 & $-$18\\

1143$-$245  & 1.9 & +72 & 2.1 & +80 & 0.7 & $-$18 & 0.7 & $-$27 & 1.3 & $-$43 & 1.4 & $-$46\\

2134$-$004  & 0.1 & & 0.1 & & 0.9 & +60 & 0.9 & +43 & 0.5& $-$4.0 & 0.5 & +0.0\\
\hline\hline

frequency &  1380 & & 1630 & & 4815 & & 4865 & & 8435 & & 8485&\\
source & $\%$ & PA & $\%$ & PA & $\%$ & PA & $\%$ & PA & $\%$ & PA & $\%$ & PA\\
\hline

 0711+356  & 1.3 & +5.0 & 1.0 & $-$33 & 1.8 & +83 & 1.8 & +80 & 1.9 & +76 & 2.0 & +79\\
\hline\hline

frequency &  1465 &&  1665 &&  4835 && 14915 && 14965 & & &\\
source    & $\%$ & PA & $\%$ & PA & $\%$ & PA & $\%$ & PA & $\%$ & PA &&\\
\hline
 
0738+313  & 0.4 & +30 & 0.3 & +24 & 6.3 & +45 & 6.6 & +26 & 6.7 & +25 &&\\
\hline
\end{tabular}
\end{center}
\vspace{6pt}\par\noindent
Note.  Columus are source name, fractional polarization in $\%$, PA (polarization position angle) in degree.
\par\noindent
The data were taken from Stanghellini et al. (1998), except for those of 0738+313 which were adapted from Saikia et al. 1998\\ 
\end{table*}

\begin{table*}
\caption{{\it F}-test of the model fit}
\begin{center}
\vspace{6pt}
\begin{tabular}{cccc}
\hline\hline
Source name & Degree of freedom & $F_{\rm rat}$ [=S(SSA)/S(FFA)]& Significance level [$\%$]\\
\hline
 0248+430 &  4 & 2.067 (0.137/0.066) & 24.97\\
 0711+356 & 4 & 8.586 (0.103/0.012) & 3.04\\
 0738+313 & 3 & 2.431 (0.073/0.030) & 24.24\\
 0743$-$006 &  2 & 1.304 (0.043/0.035) & 45.31\\
 1143$-$245 & 4 & 1.207 (0.038/0.029) & 40.15\\
 2134$-$004 & 2 & 0.983 (0.046/0.047) & 50.43\\
\hline
\end{tabular}
\end{center}
\vspace{6pt}\par\noindent
The columns are: source name, degree of freedom, $F_{\rm rat}$, and significance level for $F_{\rm rat}$[$\%$]. 
\par\noindent
\end{table*}

\begin{center}
\begin{figure}
\label{BL_LAC}
\psfig{figure=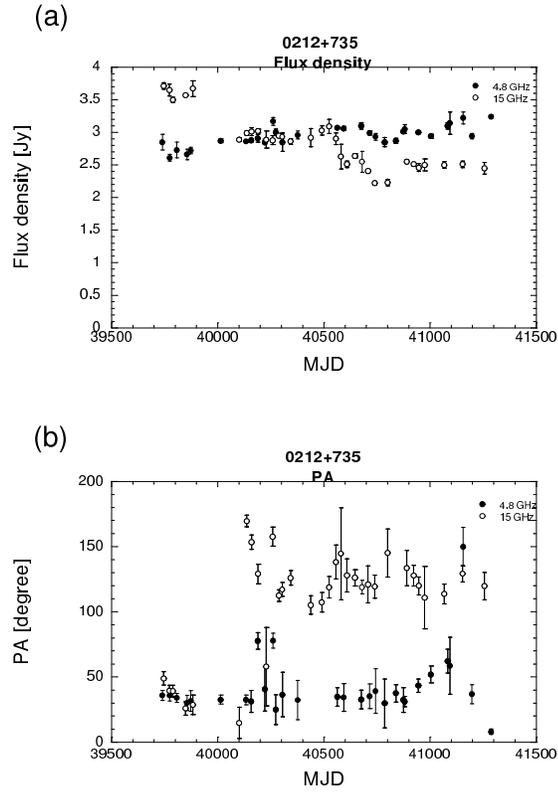,width=8.5cm}
\caption{(a) Flux density as a function of MJD. The total flux
density indicates that the emission had been optically thick
from 4.8 to 14.5~GHz before MJD~40100, and then became
optically thin at 14.5~GHz thereafter. (b) PA variation as a function
of MJD. There is a $90^{\circ}$ jump in PA at 14.5 GHz after
MJD~40100.}
\end{figure}
\end{center}

\begin{center}
\begin{figure}
\label{pol-lambda}
\psfig{figure=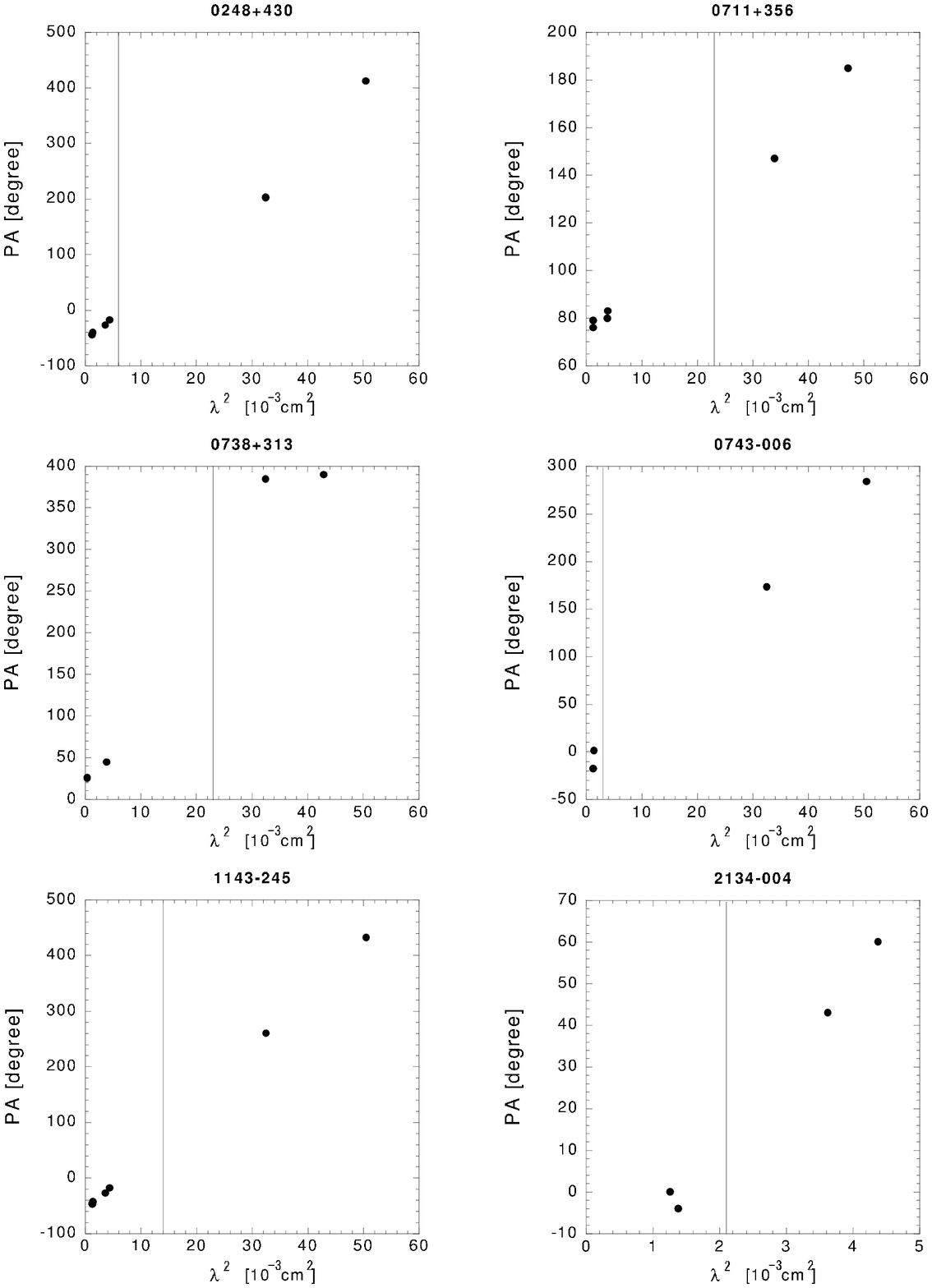,width=18cm}
\caption{PA as a function of $\lambda^2$ for the six GPS sources. There are four to six data points for each source. The vertical line shows the spectral peak wavelength.  In most cases it seems that a straight line can be fitted throughout the peak wavelength,
in support of the FFA model.}
\end{figure}
\end{center}

\begin{center}
\begin{figure}
\label{F-test}
\psfig{figure=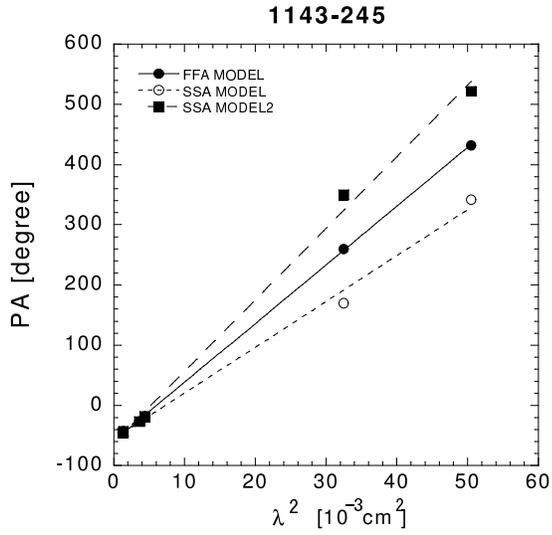,width=8.5cm}
\caption{PA as a function of $\lambda^2$ for 1143$-$245.  Two data points at longer wavelengths are also plotted with offsets of $\pm 90^{\circ}$.  Due to the high $RM$, these offset data could be fitted well by straight lines.}
\end{figure}
\end{center}

\begin{center}
\begin{figure}
\label{F-test}
\psfig{figure=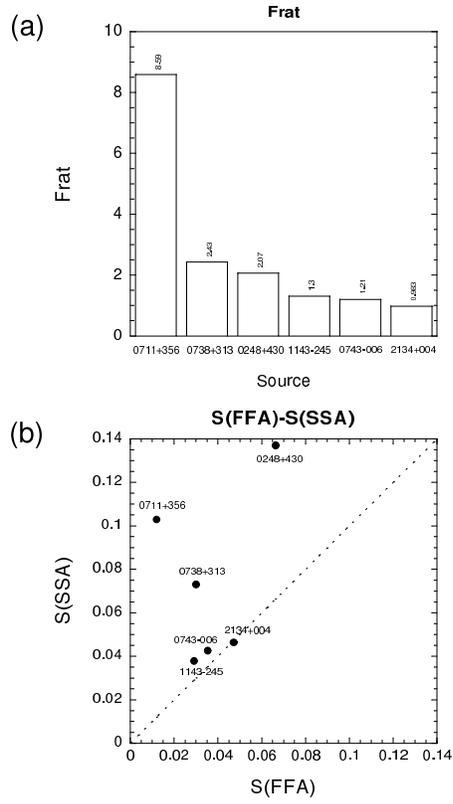,width=8.5cm}
\caption{(a) The vertical bar indicates $F_{\rm rat}$
[=S(SSA)/S(FFA)]. These are all greater than 1, except  
2134+004. (b) Plot of S(SSA) against S(FFA). The dotted line indicates
S(FFA)=S(SSA). For all sources, except 2134+004, the data points are
above the dotted line.}
\end{figure}
\end{center}

\begin{center}
\begin{figure}
\label{F-test}
\psfig{figure=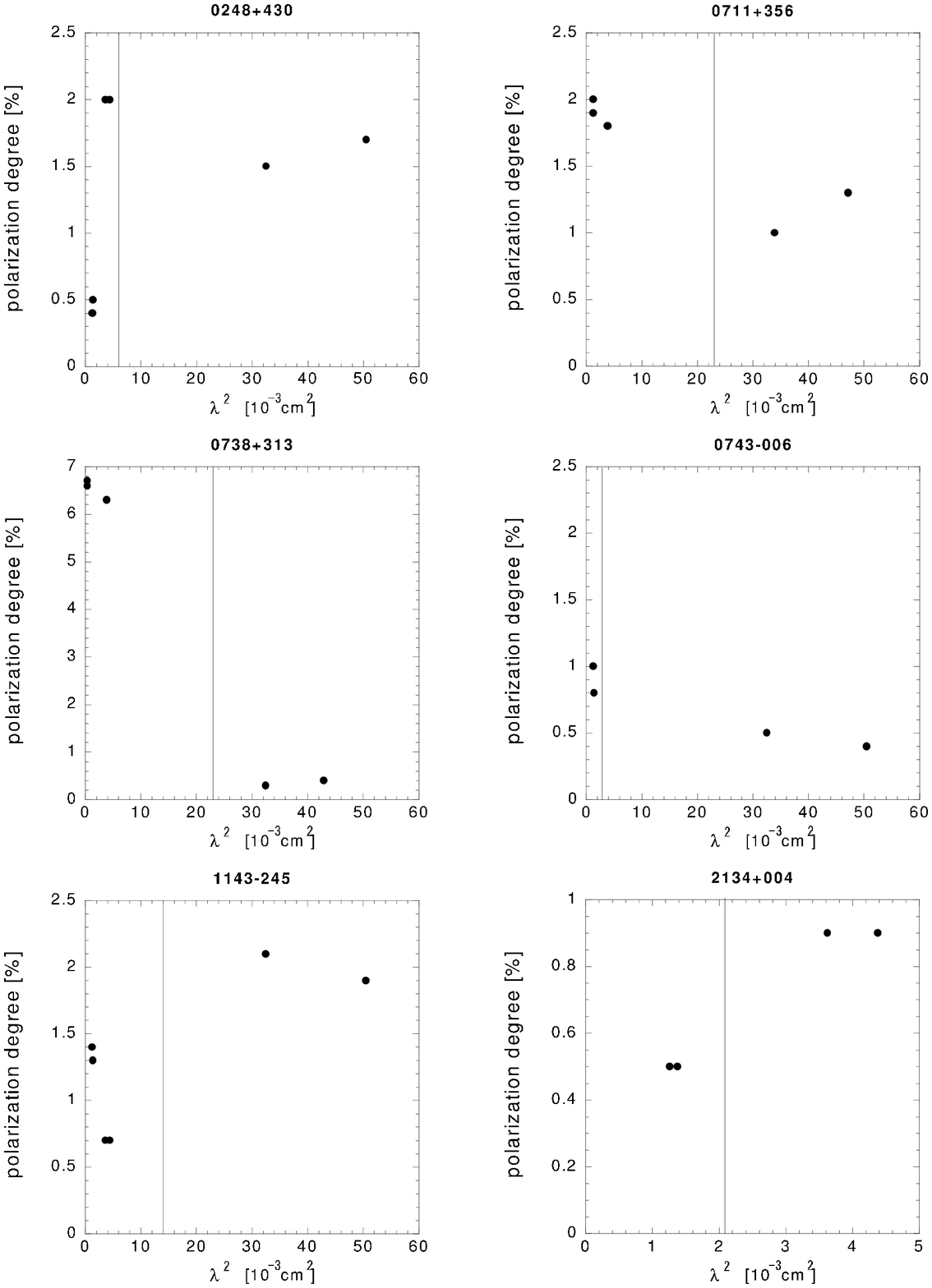,width=18cm}
\caption{Fractional polarization as a function of $\lambda^2$. The
vertical line shows the spectral peak wavelength.  All except 0738+313
show essentially the same fractional polarization in both optically
thin and thick frequencies, again in favor of FFA.}
\end{figure}
\end{center}

\end{document}